# Efficient broadband terahertz generation by above band-gap excitation of the pyroelectric ZnSnN$_2$

T. S. Seifert[1]*, H. Hempel[2], O. Gueckstock[1], R. Schneider[3], Q. Remy[1], A. Fioretti[4], T. Unold[2], S. Michaelis de Vasconcellos[3], R. Bratschitsch[3], R. Eichberger[2], K. Dörr[5], A. Zakutayev[4], T. Kampfrath[1]

[1]Freie Universiät Berlin, Berlin, Germany

[2]Helmholtz Zentrum Berlin, Berlin, Germany

[3]Universität Münster, Münster, Germany

[4] National Renewable Energy Laboratory, Golden, Colorado, United States

[5]Martin-Luther-Universität Halle-Wittenberg, Halle, Germany

*corresponding author (tom.seifert@fu-berlin.de)

**Abstract.**

Terahertz (THz) radiation is a powerful probe of low-energy excitations in all phases of matter. However, it remains a challenge to find materials that efficiently generate THz radiation in a broad range of frequencies following optical excitation. Here, we investigate a pyroelectric material, ZnSnN$_2$, and find that above-band-gap excitation results in the efficient formation of an ultrafast photocurrent generating THz radiation. The resulting THz electric field spans a frequency range from below 1 to above 30 THz. Our results suggest that the photocurrent is primarily driven by an ultrafast pyroelectric effect where the photo-excited carriers screen the spontaneous electric polarization of ZnSnN$_2$. Strong structural disorder reduces the photocarrier lifetime significantly and, thus, enables broadband operation. ZnSnN$_2$ shows similar THz-emitter performance as the best spintronic THz emitters regarding bandwidth and amplitude. Our study unveils the large potential of pyroelectric materials as efficient and broadband THz emitters with built-in bias fields.

**Introduction.**

Terahertz (THz) radiation with frequencies between 1 and 30 THz can resonantly interact with many elementary excitations, such as phonons, magnons or excitons, in solid materials [1]. Therefore, THz spectroscopy has provided numerous useful insights into the low-energy physics in diverse material systems [2, 3]. To improve THz-spectroscopy systems, the field of THz photonics constantly searches for more efficient and broadband sources of THz radiation [4]. State-of-the-art table-top THz sources rely on the down-conversion of photons from the optical to the THz spectral range. The respective process of optical rectification is typically realized by below-band-gap excitation of polar semiconductors, such as ZnTe [5], or above-band-gap excitation of electrically biased semiconductors, such as low-temperature grown GaAs [6]. In the latter devices, the electrical bias field accelerates the photo-generated mobile charge carriers, thereby inducing an ultrafast current burst that radiates a THz pulse into the far-field.

To replace the extrinsic electric bias field by internal fields, one can use THz-emitter materials with a nonvanishing permanent electric polarization $P$. An example are pyroelectrics, which include ferroelectrics whose $P$ can be switched by an external electric field. A prominent representative in this regard is the room-temperature multiferroic bismuth orthoferrite BiFeO$_3$ (BFO), in which above-band-gap excitation leads to a photocurrent that flows along the axis defined by $P$ [7]. However, typical THz-generation efficiencies from BFO [8] were reported to be much lower than for standard THz emitters like ZnTe. Other ferroelectrics show more efficient THz generation but require operation below room temperature, such as SbSI [9], or cannot be upscaled to large thicknesses, such as the van der Waals material NbOI$_2$ [10]. Previous works on the pyroelectric InN [11, 12] reported efficient THz generation between 1 and 3 THz that was tentatively associated with carrier acceleration in an electric field caused by the spontaneous polarization.

Thus, it remains a challenge to identify efficient and scalable THz generators based on in-built electric fields that operate over a wide THz frequency range at room-temperature. Solutions to this challenge bear a large application potential in THz-spectroscopy systems [13]. From a broader perspective, a better understanding of photocurrents in pyroelectrics can have a substantial impact on future energy-harvesting strategies [14].

Here, we study photo-induced THz emission from the pyroelectric semiconductor ZnSnN$_2$ (ZSN). We excite thin films of ZSN above its band gap of 1.1 eV [15] with femtosecond laser pulses (Fig. 1a) and observe the generation of ultrabroadband THz pulses covering the range from 1 to 30 THz without any spectral gaps. We find a THz generation efficiency that is about three orders of magnitude higher than that from a standard THz source like ZnTe with a similar thickness of about 1 µm. We discuss different microscopic phenomena that may explain this efficient THz-field generation. The most plausible scenario is a pump-induced polarization-screening current, i.e., an ultrafast pyroelectric effect, whose relaxation time of far below 1 ps is set by the crystallite size of <100 nm in our samples. Our study suggests a large potential of the abundant class of pyroelectric materials for the field of THz photonics.

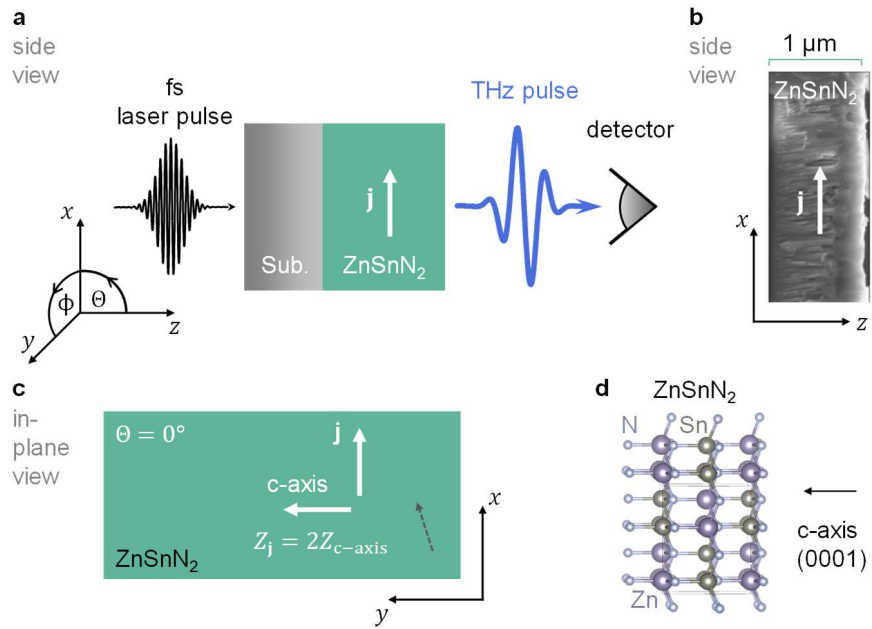

**Figure 1. Studied material and experimental setup.** a) An incident femtosecond (fs) optical pump pulse launches a photocurrent of density $j$ in the plane of a ZnSnN$_2$ thin film. Consequently, an electromagnetic pulse with terahertz (THz) frequencies is emitted into the optical far-field, where it is detected by electro-optic sampling. The ZnSnN$_2$ is grown on an Eagle glass substrate (Sub.). The angle $\Theta$ denotes the polar angle under which the sample is oriented with respect to the $z$-axis. The angle $\phi$ denotes the azimuthal angle that accounts for sample rotation about its surface normal. b) Scanning-electron micrograph of a ZnSnN$_2$ thin-film cross section. c) In-plane sample view indicating the in-plane c-axis direction (see Fig. S1) perpendicular to the direction of the photocurrent. The THz impedance is largest along $j$ and a factor of about 2 smaller along the c-axis (see main text for more details). The dashed arrow indicates the direction of the Zn/Sn compositional gradient. The $x$-$y$-$z$ coordinate system is fixed to the laboratory frame. d) Crystal structure of wurtzite ZnSnN$_2$.

## RESULTS AND DISCUSSION

**THz-emission raw data.** Fig. 2a depicts raw THz-emission signals from the 1-µm-thick ZSN sample in the stochiometric region. For comparison, the THz-emission signal from a reference sample, a trilayer spintronic THz emitter [W(2nm)|Co$_{40}$Fe$_{40}$B$_{20}$(1.8 nm)|Pt(2 nm)] on a sapphire substrate (TeraSpinTec GmbH) [2] excited under identical conditions, is also shown. A comparison to a commercial photoconductive antenna is displayed in Fig. S2. The inset of Fig. 2a reveals that the THz-emission signal from ZSN grows linearly with the incident pump fluence. It, therefore, arises from a second-order nonlinear process, analogous to any standard photocurrent.

The respective THz spectra obtained via a Fourier transformation are shown in Fig. 2b. ZSN features a remarkably broadband THz emission ranging from below 1 THz to 30 THz with an efficiency approaching that of the spintronic THz emitter. We note that the detected emitter bandwidth is currently limited by the pump- and probe-pulse duration, and that the spectral dip at around 5 THz is due to the detector response function.

Interestingly, we do not observe pronounced phonon absorption bands in the THz-emission spectrum of ZSN. Such suppression of phononic features is surprising given the polycrystalline (not amorphous) nature of our sample [16]. This observation suggests a prominent role of screening of the phonons by charge carriers, given the reported DC conductivity of about 1-10 kS/m for our samples [15].

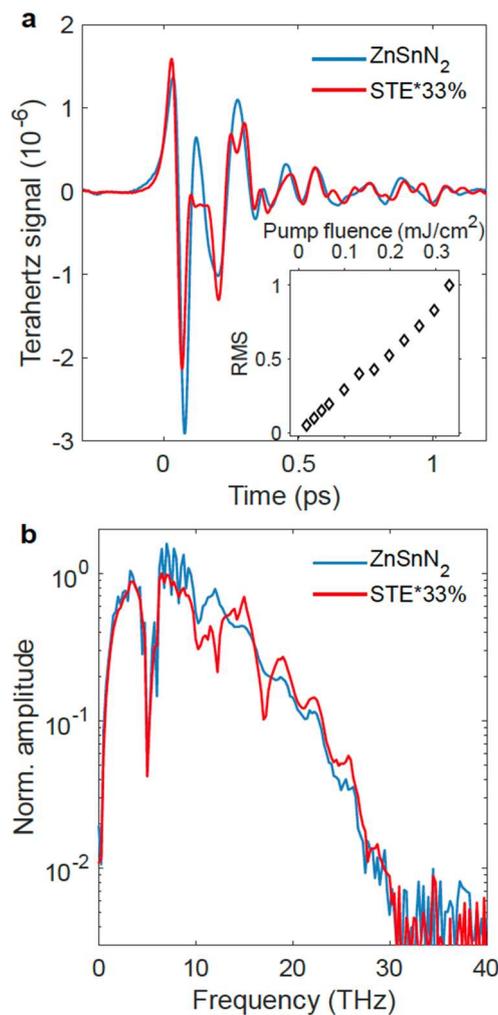

**Figure 2. Raw terahertz-emission signal from ZnSnN$_2$.** a) THz-emission waveform from ZnSnN$_2$ and, for comparison, from an optimized spintronic THz emitter (STE, divided a factor of 3), both measured with a 10 µm-thick ZnTe electro-optic detection crystal. Inset: Root mean square (RMS) of the terahertz signal from ZnSnN$_2$ vs incident pump fluence. b) Amplitude spectra of the two waveforms shown in a). Both spectra are normalized by the same factor, and the STE spectrum is subsequently divided by a factor of 3. The pronounced spectral dip at around 5 THz is related to the detector response function.

**Photocurrent characterization.** In Fig. 3a, we show THz signals for pump-photon energies larger (1.5 eV) and smaller (1.0 eV) than the electronic bandgap of ZSN (about 1.1 eV). For above-band-gap excitation, the THz-emission amplitude is strongly enhanced. This behavior contrasts with that of the STE reference sample, where the 1.5 eV and the 1.0 eV excitation lead to the same THz emission (Fig. 3a inset). The THz-emission amplitude varies by less than 20% when the polarization direction of the linearly polarized pump beam is rotated (inset of Fig. 3b). The dependence of the THz-emission signal on sample thickness is displayed in Fig. 3b. The signal amplitude increases from the 0.5 µm to the 1 µm sample and shows signatures of saturation at a thickness of 2 µm.

To summarize, optically induced THz emission from ZSN is significantly more efficient for above- than below-band-gap excitation. Further, its amplitude saturates above the pump-pulse absorption depth of about 1 µm [15] and has only a minor pump-polarization dependence. These findings indicate that THz emission from ZSN relies on deposition of pump-pulse energy in the ZSN bulk. The saturation for large sample thicknesses suggests a minor importance of pump-light absorption gradients for the photocurrent generation.

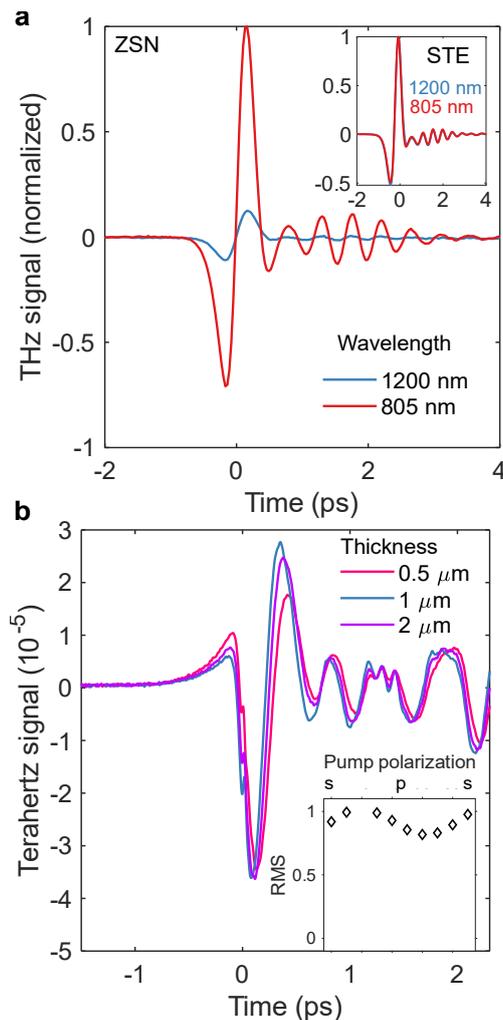

**Figure 3. ZnSnN$_2$ THz-signal symmetries: below/above band gap excitation, thickness and pump-polarization dependence.** a) THz-emission signals for excitation below (1200 nm, 1 eV) and above

(805 nm, 1.5 eV) the band gap of ZnSnN$_2$. The two waveforms are normalized by the maximum of the waveform for 805 nm pump wavelength. The inset shows the emission from the STE reference for the same pump wavelengths (same normalization as in main panel). All curves are under identical conditions. b) THz-emission signal for various ZSN thicknesses. Inset: Linear pump-polarization dependence of the root mean square (RMS) of the THz-emission signal. The RMS values have been normalized by their maximum.

**Photocurrent symmetries.** Figure 4 shows the THz-emission signal as a function of the sample's azimuthal angle $\phi$ (Fig. 4a) and polar angle $\Theta$ (Fig. 4b). In this experiment, we rotate the sample about the $z$ or $y$ axis, respectively, and detect the $x$-polarized component of the emitted THz electric field (see Fig. 1c).

We observe that the THz-emission amplitude approximately scales with the sine of the sample azimuthal angle $\phi$ about the sample normal (Fig. 4a). The question arises whether the almost complete modulation of the THz signal by $\sin\phi$ is a property of (i) the photocurrent $j$ or (ii) the conversion of the photocurrent $j$ into the THz field. As the sample normal is a principal optical axis (see XRD scans in Fig. S1) and as incident and emitted beams are along the sample normal, the THz electric field $E$ behind the sample and, thus, the electro-optic signal arises from the in-plane component $j_{IP}$ of $j$. If we rotated the sample azimuthally by $\phi$ and the pump field likewise, we would rotate $j_{IP}$ and $E$ in the same way. However, simultaneous rotation of the pump polarization is unnecessary because the THz-emission signal is mostly independent of the pump polarization direction (Fig. 3b). Therefore, rotation of the sample by $\phi$ about the sample normal simply rotates $j_{IP}$ and $E$ equally.

Note that $j_{IP}$ and $E$ are, in general, not parallel because the in-plane THz conductance of the ZSN film is highly anisotropic (Fig. S3). Importantly, such in-plane conductance anisotropy can modulate the direction and magnitude of $E$ but not its polarity (Fig. 4a). However, as in our sample the direction of smallest THz conductivity and, therefore, largest impedance coincides with the direction of $E$ (Fig. 1d), $j_{IP}$ and $E$ are indeed parallel.

In other words, the THz-emission signal originates from an in-plane charge current $j_{IP}$, the direction of which is tied to the sample structure and that is roughly perpendicular to the sample's c-axis (Fig. 1c and Fig. S1). This notion is further substantiated by our finding that the THz-signal amplitude reduces symmetrically with the polar angle $\Theta$ (Fig. 4a). An out-of-plane-oriented photocurrent would lead to an asymmetric contribution with respect to $\Theta$ instead.

In additional measurements, we turn the sample from $\Theta = 0°$ to 180° (Fig. 1a). We find a sign reversal of the THz-emission signal (see Fig. S4), supporting its electric-dipole origin.

It is interesting to note that these ZnSnN$_2$ samples have their c-axis (001 direction) lying largely in the plane of the substrate. According to XRD measurements, the preferential orientation of the <100 nm ZnSnN$_2$ grains has mostly (100) and (101) plane vectors pointing up from the substrate (Fig. S1). Therefore, the (001) c-axis plane vector is parallel to the plane of the substrate. This configuration is quite unusual for nitrides with wurtzite structure, which often tend to show strong c-axis preferential orientation with (001) plane vector perpendicular to the substrate. Further sample characterization by recording XRD pole figures indicates that the c-axis direction is nearly perpendicular to the direction of the ultrafast photocurrent (Fig. S1). This unusual preferential orientation may be related to the THz emission observed in these experiments.

To sum up, we find that ultrafast optical excitation of ZSN above its electronic band gap launches an ultrafast charge current. The latter flows in the ZSN-film plane and delivers a linearly polarized THz electric field, the direction of which is 20° relative to the compositional Zn/Sn gradient in the ZSN sample. The in-plane direction, along which the THz electric field is polarized, coincides with the direction of the largest THz impedance. It is almost perfectly perpendicular to the sample's c-axis (Fig. 1c), which, for these (100)/(101)-oriented samples, lies mostly in the plane of the substrate (Fig. S1a).

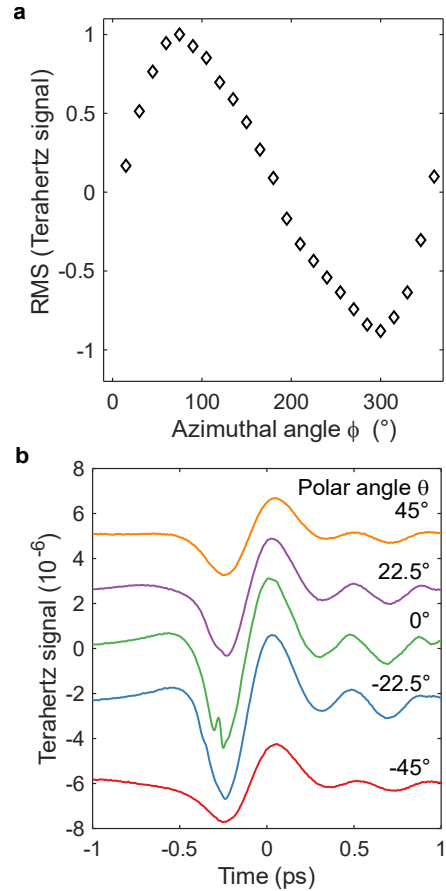

**Figure 4. Ultrafast charge-current symmetries in ZnSnN$_2$.** THz-emission signal dependence on (a) the azimuthal angle $\phi$ for $\Theta = 0°$ (see Fig. 1c) and (b) on the polar angle $\Theta$ for $\phi = 0°$ measured with a THz wire-grid polarizer behind the sample. We detect the $x$-polarized THz radiation that is perpendicular to the axis of rotation in panel b, i.e., to the $y$ axis. RMS - root mean square of the THz-emission signal (multiplied with the polarity of the respective THz waveform and normalized by its global maximum).

**Photocurrent dynamics.** Finally, we turn our attention to the ultrafast dynamics that is induced by the optical pump pulse. To this end, we extract both the ultrafast evolution of the photocurrent emitting the THz pulse (Fig. 5a) as well as the transient pump-induced changes in the THz conductance of the ZSN film (Fig. 5b). We find that both the photocurrent and the transient conductance rise immediately once the pump pulse excites the sample. Note that the experimental time resolution is about 150 fs for the measurement of the transient THz conductance and about 20 fs for the charge-current measurement. The initial relaxation of both transients happens on time scales of our respective time resolutions. These ultrashort relaxation dynamics suggest a dominant role of carrier relaxation at the nanocrystallite boundaries (Fig. 1b and Fig. S5). The long-term dynamics of the transient THz conductance show a slower evolution that might be related to carrier trapping [17].

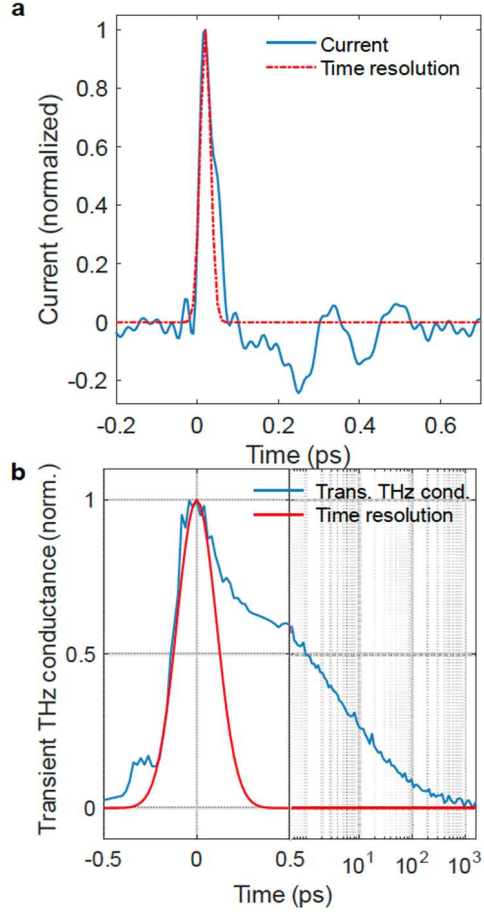

**Figure 5. Ultrafast charge-carrier dynamics in ZnSnN$_2$.** (a) Ultrafast charge current as extracted using a 50 µm ZnTe reference emitter and a 250 µm GaP electrooptic sensor [18]. The current is normalized by its maximum. (b) Transient THz sheet conductance $\Delta G_s$ (normalized by its maximum) following above-band-gap excitation. Please note the change of the time-axis scaling from linear to logarithmic scale after 0.5 ps. In panels a and b, the red curve indicates the experimental time resolution, which is given by a Gaussian with a FWHM of 20 fs and 150 fs, respectively.

**INTERPRETATION**

**Driving force.** We now discuss possible microscopic origins of the ultrafast photocurrent $j$ flowing perpendicular to the c-axis of ZSN. In other words, we look for driving forces that launch an ultrafast charge current $j$ after the pump pulse creates charge carriers by absorption above the band gap.

First, we note that we can rule out any role of pump-induced injection or shift currents because we do not observe a notable pump-helicity or linear-polarization dependence of our signal (see Fig. S6).

Second, two important driving forces for $j$ are (i) gradients in temperature, leading to Seebeck-type effects [19, 20, 21, 22], or (ii) built-in electric fields, reminiscent of the bias field in photoconductive antennas for THz-field generation [23], which drive a charge current via the ultrafast pyroelectric effect involving electronic excitation.

Option (i) is unlikely given the saturation of the THz-charge-current amplitude for film thicknesses above 1 µm, where excitation gradients are increasing (Fig. 3b). We note that this growth of temperature gradient with thickness might be masked by the concurrent decrease in impedance. For this reason, we simulate the thickness dependencies of the two driving forces, (i) and (ii) and compare it with our experimentally measured thickness dependence of the THz-emission signal (Fig. 3b). As can be seen in Fig. S7, a Seebeck-type driving force is inconsistent with our experimental results. Consequently, we exclude scenario (i) as further substantiated by our symmetry analysis (SuppMat Section 3).

Option (ii) appears more plausible since ZSN is a pyroelectric. It, in principle, permits charge currents to be driven by photoexcitation of charge carriers that subsequently screen the spontaneous polarization, i.e., an ultrafast pyroelectric effect. Regarding option (ii), we note that on the ultrafast time scales as covered by our experiment, it is typically the excited electrons that are relevant for inducing polarization dynamics, whereas phononic effects evolve on time scales $> 100$ fs. However, in the perfect ZSN crystal, the direction of the pyroelectric photocurrent is strongly tied to the crystallographic c-axis (Fig. 1d), resulting in a primary photocurrent parallel to the static polarization, i.e.,

$$j \parallel P.$$

This point is in stark contrast to our experimental observation that the direction of the photocurrent is perpendicular to the c-axis in our ZSN samples, parallel to which $P$ should be oriented in a perfect ZSN crystal.

A more detailed symmetry analysis (SuppMat Section 3 and [24]) reveals that, without additional symmetry breaking, the observed photocurrent direction is inconsistent with any of the reported cation-ordered crystal structures of ZSN [25, 26]. Our symmetry analysis also reveals that the small pump-polarization-dependent THz-emission signal (Fig. 3b) might be related to a polarization screening current along the c-axis. However, most of the THz emission stems from the screening current related to $P$ that is perpendicular to the c-axis. In the following, we will discuss several scenarios that could lead to an additional symmetry breaking to resolve the apparent contradiction regarding the direction of the photocurrent.

First, the peculiar columnar-like sample morphology (Fig. 1b) could contribute to the symmetry breaking, thereby potentially leading to electric-polarization components that do not point along the c-axis.

Second, via the piezoelectric effect in ZSN [27], strain can lead to additional electric polarization components that are not along the c-axis [28]. In this sense, lattice mismatch between the substrate and the ZSN can induce strains that alter the crystal structure, which would be in line with the inconclusive XRD scans on our samples (Fig. S1). We found additional evidence for the importance of substrate-mediated strain by observing variations in the THz-emission amplitude for different orientations of $Al_2O_3$ sapphire substrates (Fig. S8). However, these films on sapphire are relaxed rather than epitaxially strained, so the effect is relatively weak.

Third, our films are polycrystalline, on both glass and sapphire substrates, there may be grains with other orientations that lead to an in-plane current that is perpendicular to the main c-axis orientation.

Fourth, we note that samples without stochiometric gradient show identical photocurrent symmetries, which suggests that the Zs/Sn gradient is of minor importance for inducing additional $P$ components [29].

We, thus, conclude that our ZSN sample has a dominant polarization $P \parallel j$ after averaging over the pump spot size that is about 30 μm in diameter (1/e²). In this picture, the pump pulse creates a photocurrent via the ultrafast pyroelectric effect involving a polarization component $P_{str}$ that most likely arises from strain. We note that the photocurrent along the c direction is most likely suppressed by grains of opposing **P** along the (001) direction, which would be the energetically favored $P$ domain configuration.

**Charge-carrier dynamics induced by $P$ dynamics.** The question arises how the generation of photogenerated carriers leads to dynamics of $P_{\text{str}}$ that manifest as an ultrafast photocurrent $j$ in our experiments.

Because the thickness of our sample film is much smaller than the THz wavelength inside the material, we can use quasi-electrostatic arguments. According to Maxwell's first equation

$$\nabla \cdot E = Z_0 c (\rho_P + \rho_{\text{ind}}) \quad (1)$$

with the speed of light $c$, the total electric field $E$ inside the material has two sources: the charge density $\rho_P = -\nabla \cdot P$ due to the ionic polarization $P$, and the charge density $\rho_{\text{ind}}$ that is induced by $E$. Because ZSN exhibits mobile electrons, $\rho_{\text{ind}}$ predominantly arises from the displacement of electrons and, thus, partially screens the predominantly ionic $\rho_P$. In linear response, one has $\rho_P + \rho_{\text{ind}} = \rho_P/\varepsilon_e = -(\nabla \cdot P)/\varepsilon_e$, where $\varepsilon_e$ is the dielectric function of the electronic subsystem [30].

Therefore, Eq. (1) can be rewritten as

$$E = \frac{-Z_0 c}{\varepsilon_e} P, \quad (2)$$

which correctly yields $E = 0$ for the limiting case $P = 0$. Eq. (2) implies that, following excitation by the pump pulse, two distinct terms can lead to a change

$$\Delta E(t) = \frac{-Z_0 c}{\varepsilon_e}\left(\Delta P(t) - \frac{\Delta\varepsilon_e(t)}{\varepsilon_e} P\right) \quad (3)$$

in the electric field and, thus, THz emission. In Eq. (3), the first term on the right-hand side arises from a pump-induced change in the ionic charge order, i.e., ionic motion, and the second term is due to a dynamic change in the electronic screening of $\rho_P$, i.e., electron motion. Here, the change $\Delta\varepsilon_e \propto \Delta n_e$ in the electronic dielectric function scales with the pump-induced change in the density $\Delta n_e$ of mobile electrons. When writing down Eq. (3), we neglected any frequency dependence of $\varepsilon_e$ and $\Delta\varepsilon_e$ over the bandwidth covered by our experiment.

To single out, which of the two terms of Eq. (3) dominates in our experiments, we note that direct $P$ dynamics would involve the motion of nuclei. Within the simple picture of a displacive-type excitation [31], the expected dynamics of the nuclear motion should follow the twice time-integrated $\Delta n_e(t)$.

However, in the experiment, we find that $j \propto \Delta E \propto \Delta n_e$ (Fig. 5a), which excludes a dominant first term in Eq. (3). In contrast, $j \propto \Delta E \propto \Delta n_e$ is fully consistent with the second term of Eq. (3), i.e., with the excited electrons screening $P_{\text{str}}$ as the dominant ultrafast photocurrent source in ZSN.

Thus, the photocurrent dynamics is fully given by

$$j(t) \propto \left[\{\Theta(t) e^{-t/\tau_{\text{pop}}}\} * I_p\right](t), \quad (4)$$

where denotes convolution and $\Theta$ the Heaviside step function. The exponential term on the right-hand side of Eq. (4) describes the photocarrier population relaxation with a time constant $\tau_{\text{pop}}$. The convolution with the pump-pulse intensity profile $I_p$ accounts for the finite time resolution of our experiment.

Similar to photoconductive switches [23], a memory effect should exist that accounts for the transition time towards the drift regime given by the momentum relaxation time $\tau_{\text{mom}}$. However, since $\tau_{\text{mom}} \approx$ 20 fs in our samples (see Fig. S5), any memory effect is faster than our experimental time resolution.

Note that we additionally performed temperature dependent THz emission studies from ZSN (see Fig. S9) and found a pronounced decrease in THz-emission amplitude below and above room temperature. Such temperature dependence would be consistent with a critical temperature for pyroelectricity around room temperature. However, to the best of our knowledge, no reported literature values of the associated Curie temperature exist.

There are several other ternary wurtzite nitride materials that have been recently synthesized [32], and may show similar THz generating properties. In addition to ZnSnN$_2$, members of this family include ZnGeN$_2$ and ZnSiN$_2$ with 1-1-2 stoichiometry, as well as Zn$_2$(V,Nb,Ta)N$_3$, Zn$_3$(Mo,W)N$_4$ with different cation ratios, and Mg-based materials (e.g. Mg$_2$SbN$_3$, MgSnN$_2$), all with wurtzite structure. Some of these wurtzite nitrides have been also predicted to be not only pyroelectric but also ferroelectric [33], with polarization reversal by electrical stimulus, similar to (Al,Sc)N [34]. These considerations, together with the experimental discovery of THz emission from ZnSnN$_2$, provide new opportunities for design and engineering of the THz emitting materials and devices for a broad range of applications.

**CONCLUSIONS**

Our results suggest that above-band-gap excitation of pyroelectric materials can serve as a promising route toward efficient THz-field emission. Imperfections in the crystalline structure enhance the THz emission in terms of bandwidth as they lead to relaxation times of the pump-induced charge currents below 50 fs, resulting in an ultrabroadband THz-emission spectrum. The latter is similar to approaches implemented in state-of-the-art photoconductive switches that are based on defect-rich low-temperature grown GaAs, yet we here reach much shorter current-relaxation times. We foresee a significant potential for optimization of this THz emitter by material engineering, e.g., by enhancing the size of ultrafast pyroelectric currents through maximizing strains to induce even larger static polarization components via piezoelectricity.

**EXPERIMENTAL SECTION**

**Sample growth and characterization.**

Films of ZSN are grown on alkali-free borosilicate (Corning Eagle XG) glass substrates heated to 200 °C and held stationary (without rotation), resulting in a Zn/Sn composition gradient. A few additional samples were synthesized on three different orientation of $Al_2O_3$ sapphire (m-, a-, r-plane). The samples of 0.5-2.0 µm thickness controlled by the deposition time (120-480 min) were deposited by radio-frequency (RF) co-sputtering from 50-mm diameter metallic Zn and Sn targets held at 35 W and 25 W, respectively. The $N_2$ precursor gas was flown at 10 sccm through a RF plasma source held at 250 W, in the presence of Ar gas flown at 10 sccm, for the total chamber pressure of 20 mTorr.

ZSN can have a cation-disordered wurtzite crystal structure (space group 186 P63mc, Fig. 1d) or lower-symmetry cation-ordered orthorhombic structure (space group 33, Pna21) [35]. In addition, a partially-ordered structure with the space group 26, Pmc21, has been reported to have similar energy [25, 26]. The classification of the crystal structure for our samples is in principle consistent with cation-disordered wurtzite structure (Fig. S1a) due to absence of low-angle reflections , yet all candidate structures have broken inversion symmetry on the unit-cell level. The orthorhombic structure of ZSN has a calculated spontaneous polarization along the c-axis of about $1.2 \, C/m^2$ relative to a hypothetical centrosymmetric version of ZSN and features five independent piezoelectric coefficients ranging from $-0.4 \, C/m^2$ to $0.8 \, C/m^2$ [36]. We note that these coefficients are of the same order of magnitude as for the standard piezoelectric material α-Quartz [27].

Macroscopically, our ZSN films have a compositional gradient ranging from $Zn_{0.8}Sn_{1.2}N_2$ to $Zn_{1.2}Sn_{0.8}N_2$ over about 5 cm along an in-plane direction (Fig. 1c) which is 70° off the $c$ direction (see Fig. 1c). The Zn-rich samples have lower carrier density and lower mobility, whereas Zn-poor samples have higher carrier density and higher mobility, as described in our previous publications, which may influence THz generation efficiency similar to InN [11, 12, 37]. Unless stated otherwise, all THz emission data is acquired under ambient conditions from the stochiometric $ZnSnN_2$ sample region with a thickness of 1 µm. On a mesoscopic scale, the ZSN thin films feature a columnar grain structure with a column diameter of less than 100 nm (Fig. 1b, Fig. S1b) and a variety of crystallographic orientations as indicated by polycrystalline XRD peaks (Fig. S1, Fig. S1a). Our X-ray diffraction studies (see Fig S1) further reveal that the c-axis of our ZSN samples (001-axis) preferentially lies in the sample plane and is perpendicular to the $j$ direction (Fig. 1c). The column length extends throughout the entire bulk of the ZSN film, i.e., along the $z$ direction of the sample (Fig. 1a, Fig. S10b). Previous studies on comparable samples revealed that the columnar grains appear slightly tilted into the direction of the Zn-Sn compositional gradient and that nanometer-scale voids or amorphous regions are present in between the columnar grains [15].

ZSN is a semiconductor with a theoretical predicted direct electronic band gap of 1.5 eV for the cation-ordered orthorhombic crystal structure [15], and the optical band gap (defined as a strong optical absorption onset of about 1.1 eV (Fig. S10a) indicated by black color (Fig. S10b), in the cation-disordered wurtzite structure of our samples [15]. In this manuscript we refer to this 1.1 eV optical absorption onset as the optical "band gap" of the material. $ZnSnN_2$ recently attracted interest due to its potential for solar-energy conversion applications [38]. This potential application provides an additional motivation to understand the generation and relaxation of photocurrents in this material, in particular on the femtosecond time scale, which is the native time scale of the relaxation of electron momentum and energy.

**THz transmission and emission measurements**

We measure the static sample transmission at THz frequencies by generating broadband THz pulses with a photoconductive switch and detect these pulses behind the sample using electro-optic sampling in a ZnTe crystal (1 mm thickness) [5]. We find an anisotropic sample response under normal incidence that yields a THz transmission and, therefore, impedance $Z$ that is twice as high along the c-axis, compared to the orthogonal in-plane direction, i.e., the $j$ direction (see Figs. 1d and S2).

The transient THz transmission is measured with an optical-pump terahertz-probe setup based on laser pulses with a repetition rate of 150 kHz, pulse duration of 150 fs, and a central wavelength of 400 nm. Photoexcitation generates a carrier sheet concentration $\Delta n_\mathrm{e}$ of 5.5×10$^{13}$ cm$^{-2}$ per pulse as derived from the absorbed power inside the sample. The THz pulse is generated by optical rectification in a ZnTe crystal and detected by electro-optical sampling in another ZnTe crystal (both 1 mm thick). The THz transmission without excitation $T$ and the photoinduced change in THz transmission $\Delta T$ are measured with a lock-in amplifier. From this measurement, we obtain the transient THz sheet photoconductance $\Delta G_\mathrm{s}$ by [39]

$$\Delta G_\mathrm{s} = \frac{-1 + n_\mathrm{sub}}{Z_0} \frac{\Delta T}{T} \qquad (4)$$

Here, $Z_0 \approx 377\,\Omega$ is the vacuum impedance, and $n_\mathrm{sub}$ the refractive index of the Eagle XG glass substrate at 1 THz ($n_\mathrm{sub} = 2$).

The transient sheet conductance is measured by scanning the delay between the pump pulse and THz pulse with a delay stage. The sum mobility spectrum $\mu_\Sigma = -\Delta G_\mathrm{s}/(ed\Delta n_\mathrm{e})$ is measured by scanning the delay between the THz pulse and the THz sampling pulse with another delay stage, and Fourier transforming the obtained THz pulses (Fig. S5).

As shown in Fig. 1c, we employ THz-emission spectroscopy to measure photocurrents with femtosecond resolution [40]. The pump pulse excites the sample. In the first setup, we use pump pulses with a nominal duration of 10 fs, incident energy of 3 nJ, an incident excitation fluence of 0.3 mJ/cm$^2$, a repetition rate of 80 MHz and a center wavelength of 800 nm. In the second setup, two different pump pulses with a nominal duration of 60 fs (35 fs), 40 µJ, 0.05 mJ/cm$^2$, 1 kHz and a center wavelength of 805 nm (1200 nm) are used to probe the samples above and below its band gap.

The pump-induced photocurrent with density $j$ emits an electromagnetic pulse with frequencies reaching the THz range. In the optical far-field, the THz electric field is detected by electro-optic sampling [5] in a ZnTe crystal with a thickness of 1 mm or 10 µm. The electro-optic probe pulse is split off from the pump pulse. The THz-induced probe ellipticity is analyzed using a quarter wave plate and a balanced detection scheme. The detection unit is sensitive to the THz electric field in the $x$-direction (see Fig. 1a).

**Acknowledgments.**

The authors acknowledge funding by the German Research Foundation through the collaborative research center SFB TRR 227 "Ultrafast spin dynamics" (project ID 328545488, projects A05 and B02) and the priority program SPP2314 "INTEREST" (projectID GE 3288 2-1, project ITISA). This work was authored in part at the National Renewable Energy Laboratory (NREL), for the U.S. Department of Energy (DOE) under Contract No. DE-AC36-08GO28308. Funding for nitride synthesis was provided by the Office of Science (SC), Basic Energy Sciences (BES). The views expressed in the article do not necessarily represent the views of the DOE or the U.S. Government.

**Conflict of Interest.**

T.S.S. and T.K. are shareholders of TeraSpinTec GmbH, and T.S.S. is an employee of TeraSpinTec GmbH. The authors declare that they have no other competing interests.

**References.**

[1] Kampfrath, T., K. Tanaka, and K.A. Nelson Nature Photonics, 2013. **7**: p. 680.
[2] Dexheimer, S.L. 2017: CRC press.
[3] Song, H.-J. and T. Nagatsuma. 2015: CRC press.
[4] Seifert, T.S., L. Cheng, Z.X. Wei, et al. Applied Physics Letters, 2022. **120**.
[5] Leitenstorfer, A., S. Hunsche, J. Shah, et al. Applied Physics Letters, 1999. **74**: p. 1516.
[6] Brown, E.R., K.A. Mcintosh, K.B. Nichols, et al. Applied Physics Letters, 1995. **66**: p. 285.
[7] Takahashi, K., N. Kida, and M. Tonouchi Physical review letters, 2006. **96**: p. 117402.
[8] Guzelturk, B., A.B. Mei, L. Zhang, et al. Nano letters, 2019. **20**: p. 145.
[9] Sotome, M., M. Nakamura, J. Fujioka, et al. Proc Natl Acad Sci U S A, 2019. **116**: p. 1929.
[10] Handa, T., C.Y. Huang, Y. Li, et al. Nat Mater, 2025: p. 1.
[11] Ascázubi, R., I. Wilke, K. Denniston, et al. Applied physics letters, 2004. **84**: p. 4810.
[12] Wang, X.Q., G.Z. Zhao, Q. Zhang, et al. Applied Physics Letters, 2010. **96**.
[13] Koch, M., D.M. Mittleman, J. Ornik, et al. Nature Reviews Methods Primers, 2023. **3**: p. 48.
[14] Tan, L.Z., F. Zheng, S.M. Young, et al. Npj Computational Materials, 2016. **2**: p. 1.
[15] Fioretti, A.N., A. Zakutayev, H. Moutinho, et al. Journal of Materials Chemistry C, 2015. **3**: p. 11017.
[16] Hendry, E., F. Wang, J. Shan, et al. Physical Review B, 2004. **69**: p. 081101.
[17] Ye, F., Z.C. Zhao, C.S. He, et al. Applied Physics Letters, 2024. **125**.
[18] Seifert, T.S., S. Jaiswal, J. Barker, et al. Nat Commun, 2018. **9**: p. 2899.
[19] Takahashi, K., T. Kanno, A. Sakai, et al. Advanced Optical Materials, 2014. **2**: p. 428.
[20] Takahashi, K., T. Kanno, A. Sakai, et al. Physical Review B, 2015. **92**: p. 094307.
[21] Yordanov, P., T. Priessnitz, M.J. Kim, et al. Adv Mater, 2023. **35**: p. e2305622.
[22] Zhang, S., Y.W. Cui, S.J. Wang, et al. Advanced Photonics, 2023. **5**: p. 056006.
[23] Auston, D.H. Applied Physics Letters, 1975. **26**: p. 101.
[24] Birss, R.R. 1964.
[25] Quayle, P.C., E.W. Blanton, A. Punya, et al. Physical Review B, 2015. **91**: p. 205207.
[26] Lany, S., A.N. Fioretti, P.P. Zawadzki, et al. Physical Review Materials, 2017. **1**: p. 035401.
[27] Ogi, H., T. Ohmori, N. Nakamura, et al. Journal of Applied Physics, 2006. **100**.
[28] Romanov, A.E., T.J. Baker, S. Nakamura, et al. Journal of Applied Physics, 2006. **100**.
[29] Huang, F., C. Hu, H. Tian, et al. Crystal Growth & Design, 2019. **19**: p. 5362.
[30] Ashcroft, N.W. and N. Mermin Physics (New York: Holt, Rinehart and Winston) Appendix C, 1976. **1**.
[31] Zeiger, H.J., J. Vidal, T.K. Cheng, et al. Phys Rev B Condens Matter, 1992. **45**: p. 768.
[32] Zakutayev, A., S.R. Bauers, and S. Lany Chemistry of Materials, 2022. **34**: p. 1418.
[33] Lee, C.-W., N.U. Din, K. Yazawa, et al. Matter, 2024. **7**: p. 1644.

[34] Fichtner, S., N. Wolff, F. Lofink, et al. Journal of Applied Physics, 2019. **125**.
[35] Lahourcade, L., N.C. Coronel, K.T. Delaney, et al. Adv Mater, 2013. **25**: p. 2562.
[36] Adamski, N.L., D. Wickramaratne, and C.G. van de Walle Journal of Materials Chemistry C, 2020. **8**: p. 7890.
[37] Fioretti, A.N., A. Stokes, M.R. Young, et al. Advanced Electronic Materials, 2017. **3**.
[38] Khan, I.S., K.N. Heinselman, and A. Zakutayev Journal of Physics-Energy, 2020. **2**: p. 032007.
[39] Hempel, H. 2020: Freie Universitaet Berlin (Germany).
[40] Braun, L., G. Mussler, A. Hruban, et al. Nature Communications, 2016. **7**: p. 13259.
[41] Haase, A., C. Genzel, D. Dantz, et al., in *Applied Crystallography*. 2001, World Scientific. p. 97.
[42] Passler, N.C. and A. Paarmann Journal of the Optical Society of America B-Optical Physics, 2017. **34**: p. 2128.
[43] Deng, F., H. Cao, L. Liang, et al. Opt Lett, 2015. **40**: p. 1282.

# SUPPLEMENTARY MATERIAL

1. **X-ray diffraction, pole figures and scanning electron micrographs**

Structural and morphological properties of the Zn-Sn-N samples are summarized in Fig. S1. The ZnSnN2 grains are preferentially oriented in (100) and (101) directions, with the usual (001) oriented grains of the wurtzite nitrides being a minor fraction according to XRD measurement with a 2D area detector (Fig. S1a). The grains are less than 100 nm in size and extend through the film thickness, according to crossectional SEM imaging (Fig. S1b).
Angle-dispersive XRD measurements were carried out by means of a 5-circle diffractometer (SEIFERT, special design) using Cu Kα radiation. More details about the diffractometer can be found in Ref. [41]. In Fig. S1c, we present the pole figure for the 001 diffraction peak revealing an orientation of the samples c axis preferentially perpendicular to the measured direction of photocurrent (see Fig. 1c).

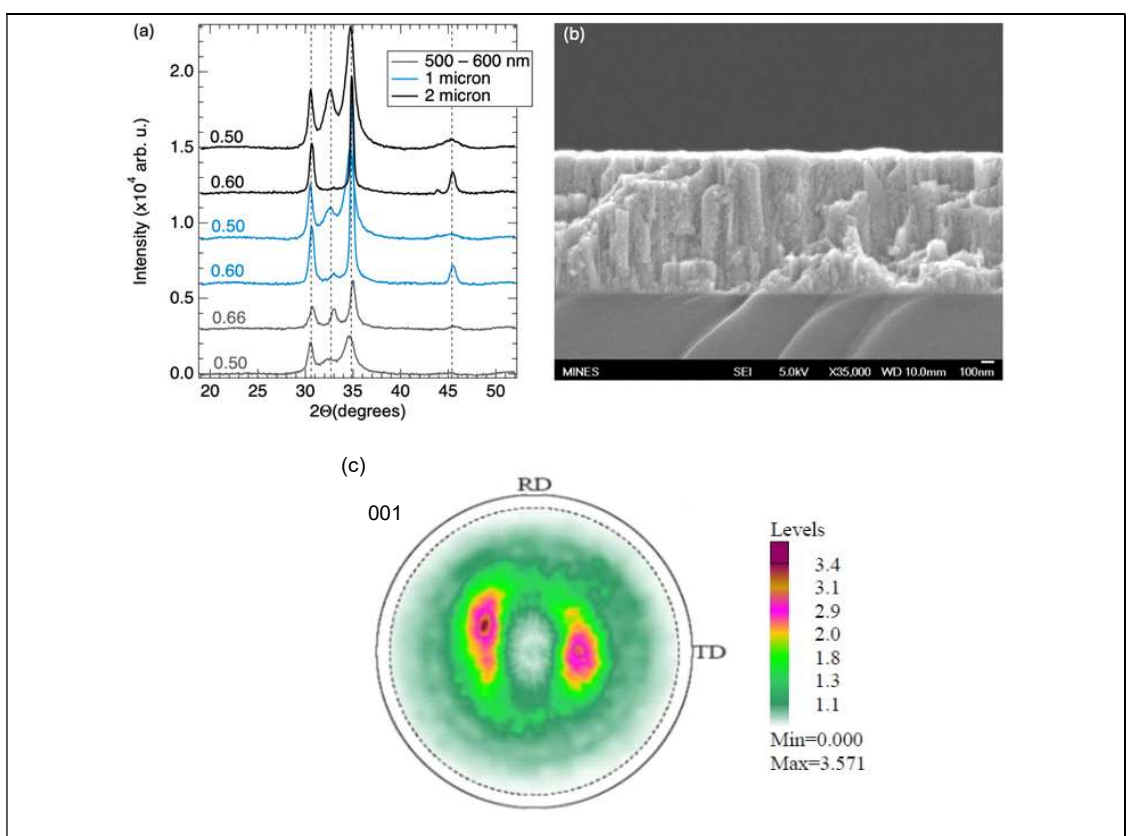

**Figure S1 | Structural and morphological properties of the ZSN sample. a** XRD patters of stoichiometric (x=0.5) and Zn-rich (x=0.6) regions of the ZnSnN$_2$ samples of the 0.5 μm, 1 μm, 2 μm thick samples, indicating mixed 100/101 preferential orientation. **b** Crossectional scanning electron microscopy image of the 1 μm thick sample, indicating columnar grain structure with <100nm grain diameter. **c** XRD Pole figure for the 001 diffraction peak (c axis) from the stochiometric sample region. The RD direction corresponds to the in-plane direction along which the photocurrent is flowing Fig. 1c. The TD direction corresponds to the in-plane direction that is perpendicular to the photocurrent.

## 2. Comparison of thickness dependencies of temperature gradients and dynamics of the electric polarization

To compare the thickness dependencies of the two driving forces, (i) gradient in temperature or (ii) pump-induced $P$ dynamics, we first calculate the electric field profile of the pump beam inside the sample [42]. We assume a refractive index of 2.3+0.1i for ZSN [43] and of 2 for the substrate.

For scenario (i), the driving force and, thus, the photocurrent amplitude should scale as

$$j_c^{\text{SSE}} \propto T(d) - T(0) \tag{S1}$$

with the electronic temperature $T$ and the ZSN film thickness $d$. The electronic temperature right after excitation is given by the absorbed pump energy, which is directly proportional to the electric field squared, i.e., to $E^2$. Thus, we find $j_c^{\text{SSE}} \propto E^2(d) - E^2(0)$.

In scenario (ii), the resulting driving force scales with the total absorbed pump energy that we can directly calculate for a given film thickness [42].

To compare these calculated values to the experiment, we additionally need to consider the impedance of the ZSN film that mediates the charge-current-to-electric-field conversion. It is calculated from

$$Z = \frac{Z_0}{1 + n_{\text{sub}} + Z_0 d \sigma_{\text{ZSN}}} \tag{S2}$$

with the free-space impedance $Z_0 = 377\,\Omega$, substrate refractvie index $n_{\text{sub}} = 2$, film thickness $d = 1\,\mu\text{m}$ and the DC conductivity of ZSN $\sigma_{\text{ZSN}} = 2 \cdot 10^3\,\text{S/m}$ [15]. Note that we neglect any frequency dependence of $\sigma_{\text{ZSN}}$ over the bandwidth covered by our experiment due to the short momentum relaxation time $\tau = 20\,\text{fs}$.

As can be seen in Fig. S7, scenario (ii) agrees well with the measured data, which means that the photocurrents are primarily driven by dynamics related to $P$.

### 3. Symmetry analysis

In order to discriminate potential mechanisms that explain THz emission from ZSN, we conducted a symmetry analysis [24]. Based on the linearity of the THz signal with the pump fluence (Fig. 2a), the photocurrent must be proportional to either $EE^*$ or to the generated temperature (or carrier density) gradient $\nabla T$. Note that in this section $E$ denotes the pump electric field. Based on the reversal of the THz-signal polarity upon reversing the sample (Fig. S4), the photocurrent can be, to lowest order, written as:

$$j_i = \chi_{iz}(\nabla T)_z \tag{S3}$$

or

$$j_i = \chi_{ijkl} E_j E_k^* P_l \tag{S4}$$

where indices $i, j, k$ and $l$ follow Einstein summation convention, and the temperature gradient is taken along the $z$ direction (Fig. 1a). The first equation corresponds to a Seebeck effect while the second one corresponds to an effect related to the spontaneous electric polarization $P$. We work in a coordinate system defined in Fig. 1a. That way, the c axis is along the $y$ direction and the observed photocurrent is along the $x$ direction. The non-zero tensor elements tensor of $\chi$ are found following the method of reference [24]. We consider the point groups $C_{6v}$ (wurtzite structure) as well as $D_2$, $C_{2v}$ and $D_{2h}$ (orthorhombic structure).

In the first case of a Seebeck effect (Eq. S3), we find that symmetry prohibits all kinds of in-plane photocurrents ($j_x = j_y = 0$). Those arguments are based on the symmetry of the bulk. We can however consider interfacial Seebeck effects by removing the symmetry operations forbidden by the presence of the interface. In this case, we find that a photocurrent along the c-axis is allowed ($j_y = -\chi_{yz}(\nabla T)_z$), only for $C_{2v}$ and $C_{6v}$ point groups, but a photocurrent along the $x$ axis is still forbidden. Thus, the symmetry analysis allows us to conclude that a Seebeck effect is not consistent with the presence of a photocurrent orthogonal to the c-axis in crystals with wurtzite and orthorhombic symmetries.

Next, we study the photocurrent related to $P$ (Eq. S4). Denoting the electric field component along the $y$ direction by $E_s$ and the one along the $x$ direction by $E_p e^{i\delta}$ (with $\delta$ a phase shift allowing the description of circularly polarized light), and fixing $P$ along the $y$ direction, we find:

$$j_y = \left(\chi_{yxxy} E_p^2 + \chi_{yyyy} E_s^2\right) P_y \tag{S5}$$

$$j_x = \left(\chi_{xxyy'} e^{i\delta} + \chi_{xyxy} e^{-i\delta}\right) E_p E_s P_y \tag{S6}$$

Only the real part of these currents has a physical significance. If instead $P$ lies along the $x$ direction, we obtain the same results together with the substitutions $x \leftrightarrow y$ and $\delta \to -\delta$. These results are valid for all four considered point groups. We note that Eq. S5 entails a pump polarizatioon independent current, whereas Eq. S6 is pump polarization dependent.

Thus, we find that only an orientation of $P$ along the $x$ direction, i.e., orthogonal to the c-axis of ZSN, is consistent with a pump-polarization independent photocurrent along the $x$ direction.

If $P$ is along the $y$ direction instead, i.e., parallel to the c-axis, only a polarization dependent photocurrent can be generated along the $x$ direction. Depending on the value of the complex numbers $\chi_{xxyy}$ and $\chi_{xyxy}$, the photocurrent can depend on the helicity of the pump. In particular, if $\chi_{xxyy}$ and $\chi_{xyxy}$ are complex conjugates of each other, the pump-polarization dependence is exactly the one observed in Figs. 3b and S6, i.e. without helicity dependence and maximum when the angle between the pump polarization and the c-axis is 45°.

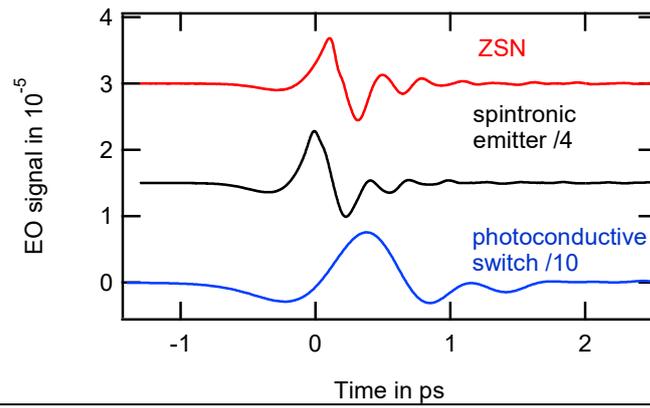

**Figure S2 | Comparison to standard THz emitters.** Raw THz emission signals from ZSN, a spintronic THz emitter (W|CoFeB|Pt) and a commercially available photoconductive switch (TeraBlast) detected with a 1 mm ZnTe crystal.

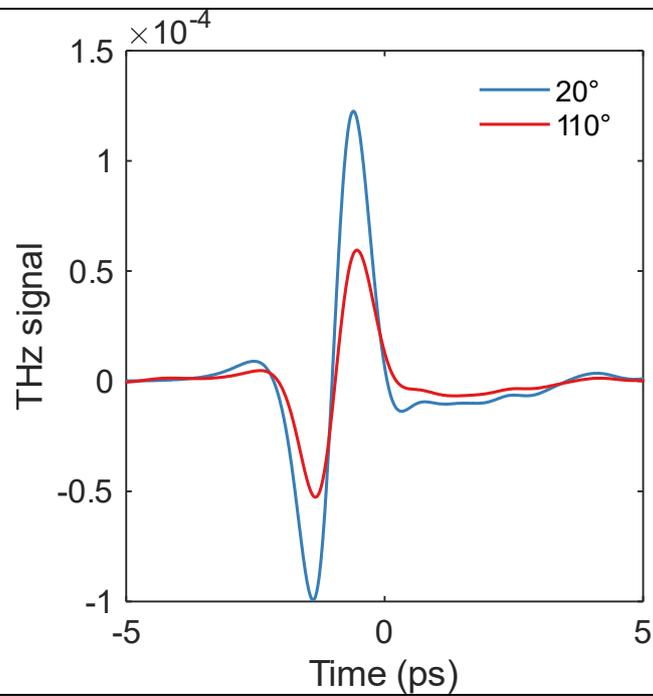

**Figure S3 | Static THz transmission.** THz signals detected after transmission through the ZSN sample for two different azimuthal orientations (c.f. Fig. 1d).

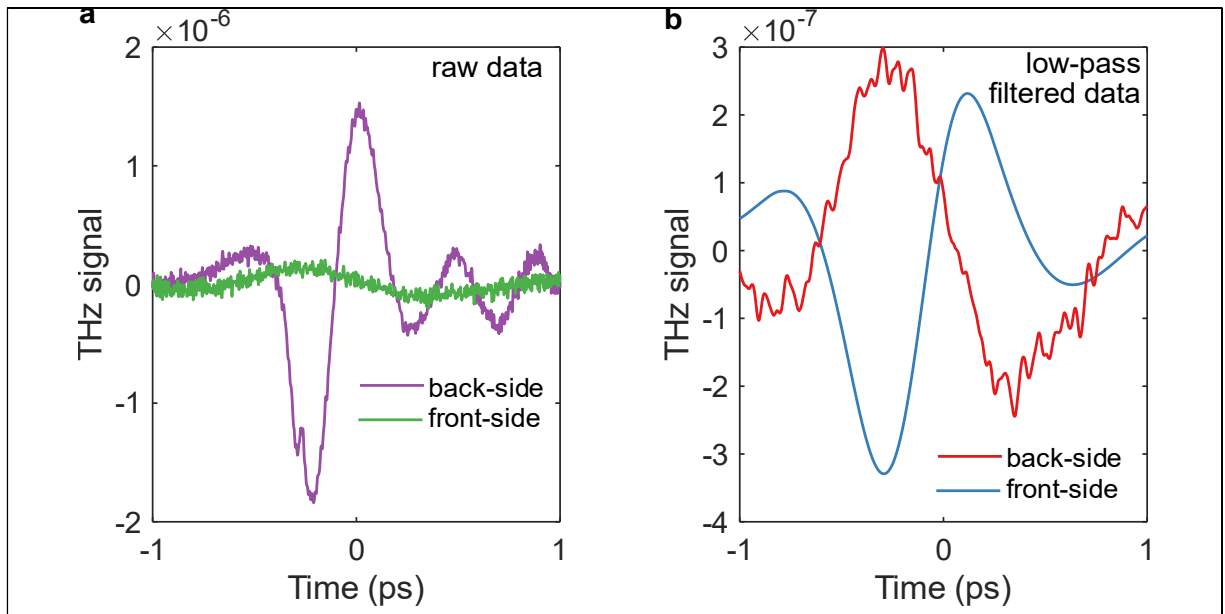

**Figure S4 | THz emission front-side vs back-side pumping. a** Raw THz emission signals from ZSN for different sample orientations. The sample is pumped under normal incidence, for two cases: ZSN|substrate (front-side) and substrate|ZSN (back-side) orientation. Waveforms are shifted horizontally to overlap. **b** Low-pass filtered data from panel a highlighting the reversed polarity for the two orientations.

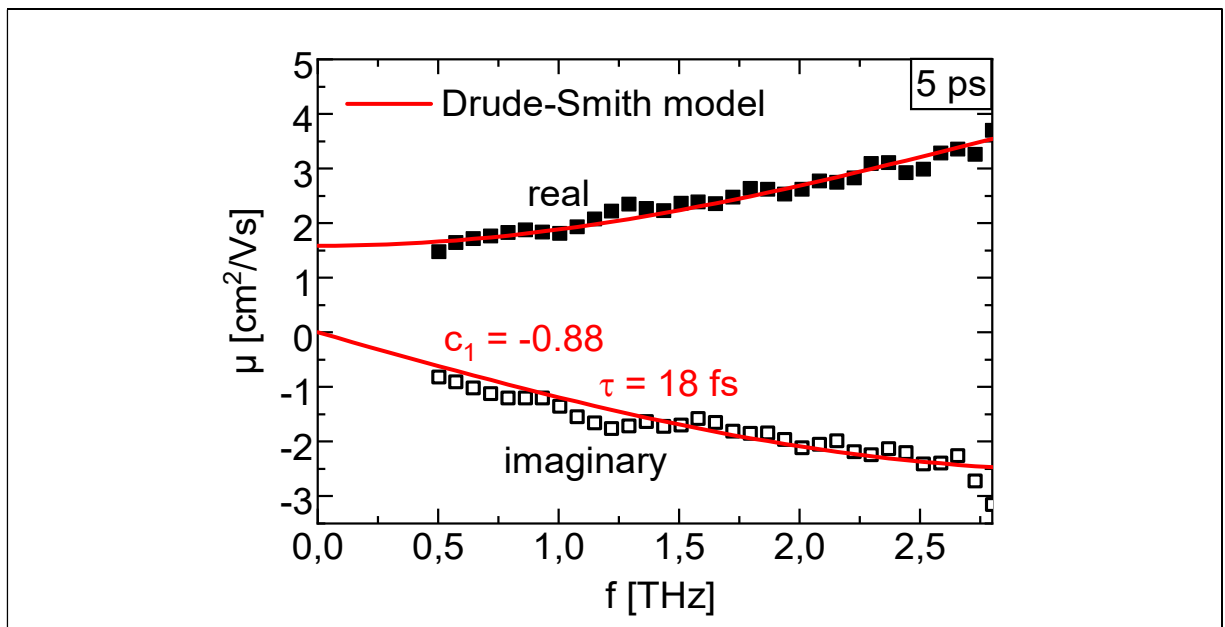

**Figure S5 | Drude-Smith fit of transient THz mobility after 5 ps.** The mobility spectrum is described by the Drude-Smith model, which reveals strong back scattering ($c_1 = -1$ for total backscattering and $c_1 = 0$ for free transport) of photogenerated charge carriers within a scattering time $\tau$ of 18 fs. Likely, this backscattering causes the relaxation of the THz-generating currents, which is found to be on the same timescale.

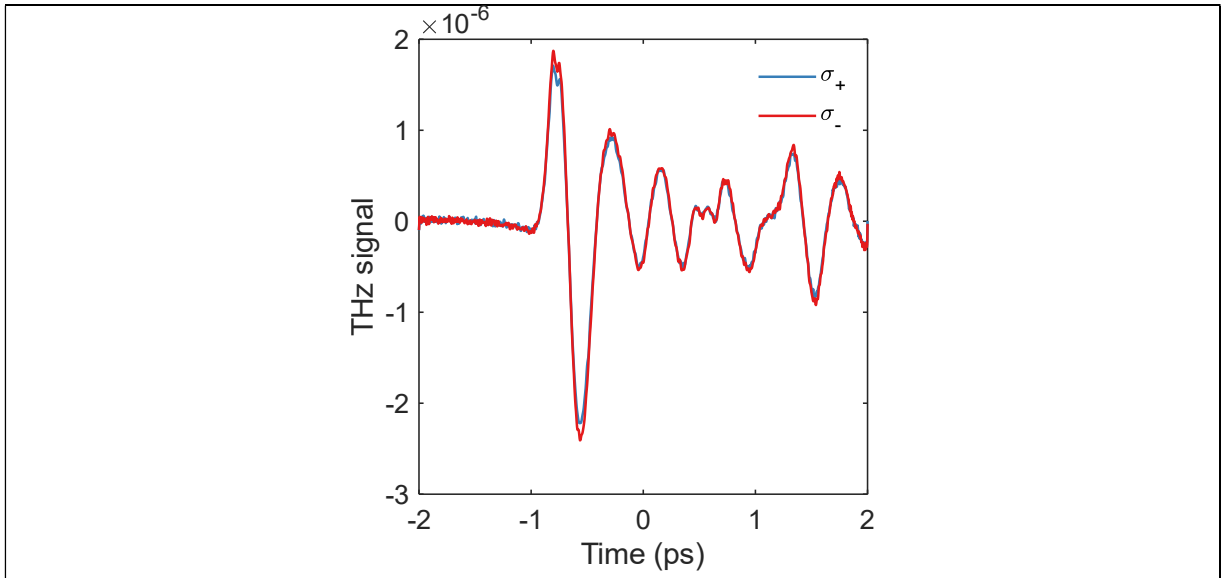

**Figure S6. Helicity dependence.** THz emission signal for left- vs. right-handed ($\sigma_+$, $\sigma_-$) excitation of ZSN.

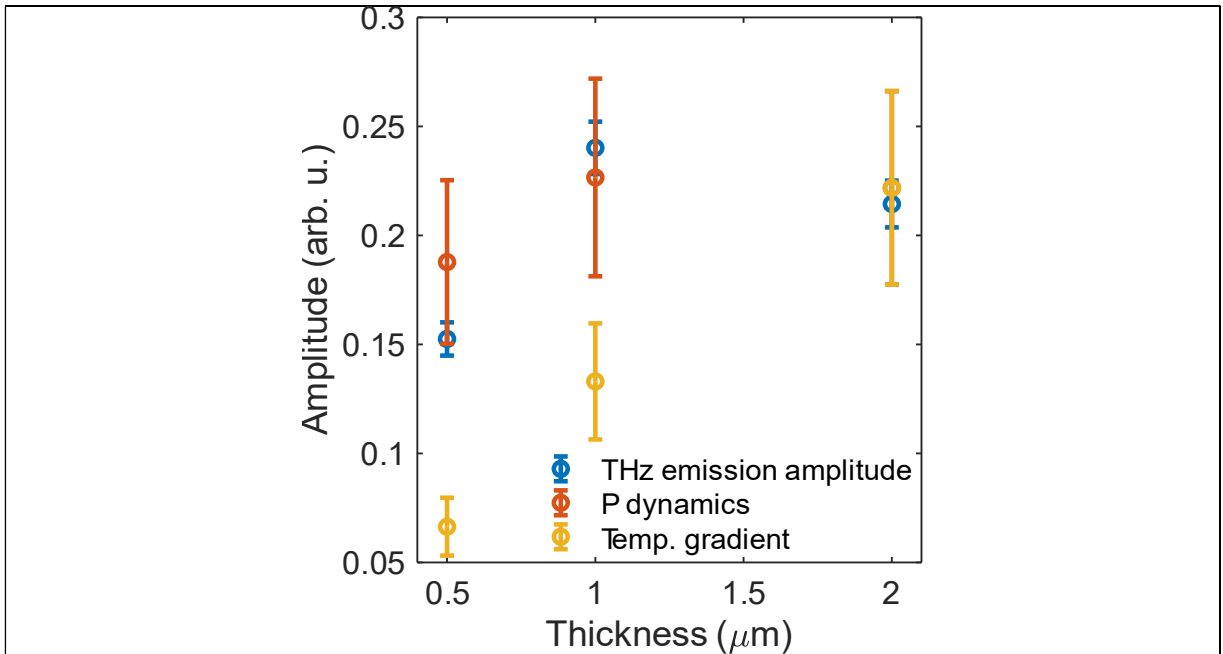

**Figure S7 | Simulations of different photocurrent driving forces.** Normalized THz emission raw-data amplitudes (blue), calculated thickness dependence of the photocurrents driven by dynamics related to $P$ (red), and of the temperature-gradient-driven photocurrents (orange). Scaling factors for each curve are chosen such that a qualitative comparison between the three thickness dependencies becomes possible. For 2 µm, the data and error bars photocurrents driven by dynamics related to $P$ and by temperature gradients exactly overlap. Error bars are chosen as 5% for the THz emission amplitude to account for the experimental uncertainties and as 20% for the two simulated thickness dependencies to account for uncertainties in the refractive index and conductivity of ZSN.

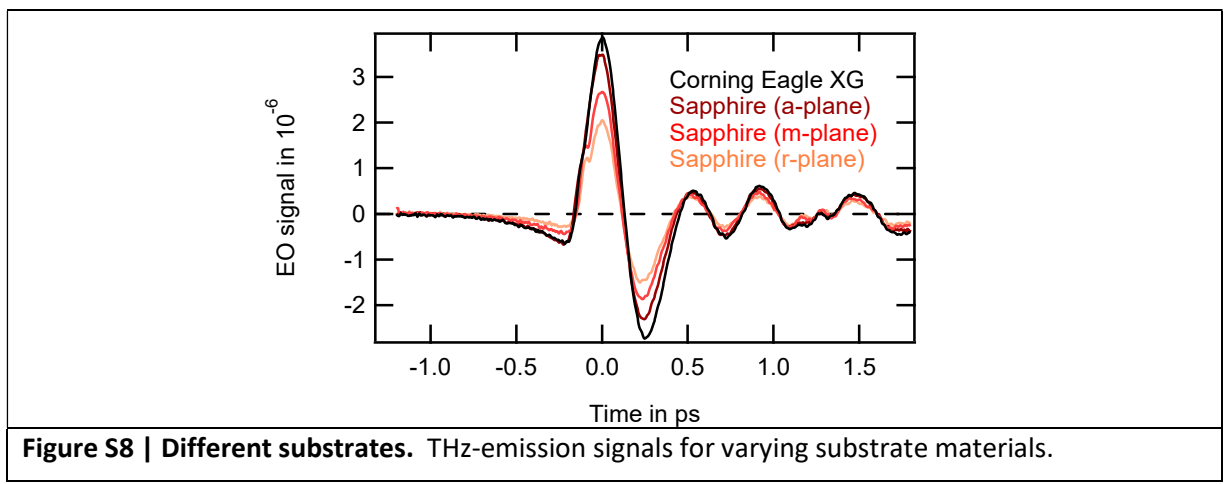

**Figure S8 | Different substrates.** THz-emission signals for varying substrate materials.

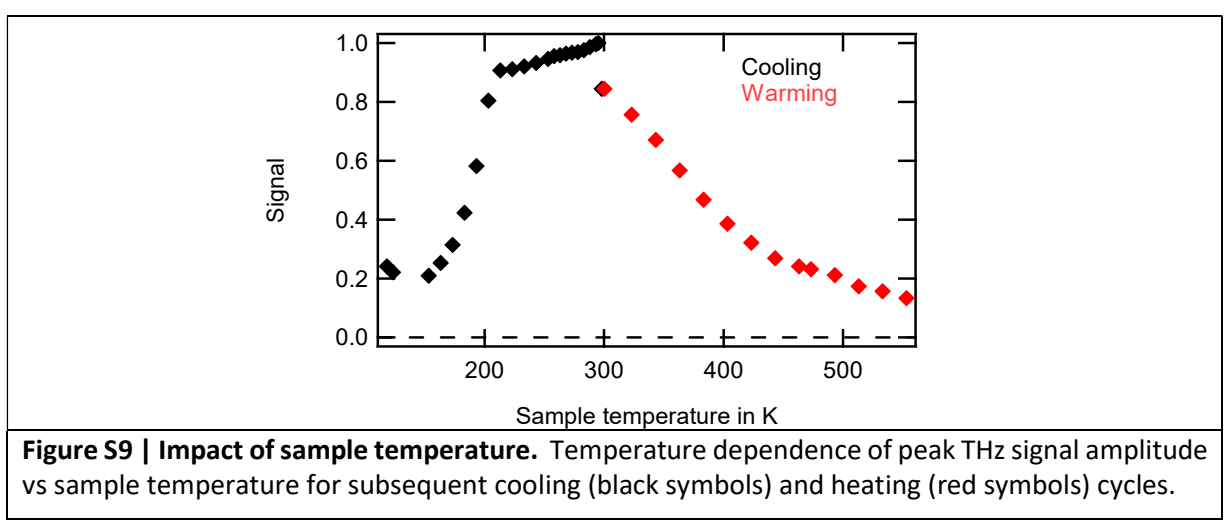

**Figure S9 | Impact of sample temperature.** Temperature dependence of peak THz signal amplitude vs sample temperature for subsequent cooling (black symbols) and heating (red symbols) cycles.

4. **ZSN absorption**

The ZnSnN$_2$ samples are heavily doped semiconductors according to optical absorption measurements shown in Fig. S10. The absorption onset in the Zn-rich x=0.6 samples is 1.-1.1 eV, and it's shifted by free carrier absorption to 1.2-1.3 eV for the stoichiometric x=0.5 samples (Fig. S10a). This indicates that the 805 nm excitation leads to band-to-band transitions with subsequent THz emission in all the measured samples. The samples deposited on sapphire Al$_2$O$_3$ substrates have dark light-absorbing appearance, consistent with these relatively optical absorption onset measurement results.

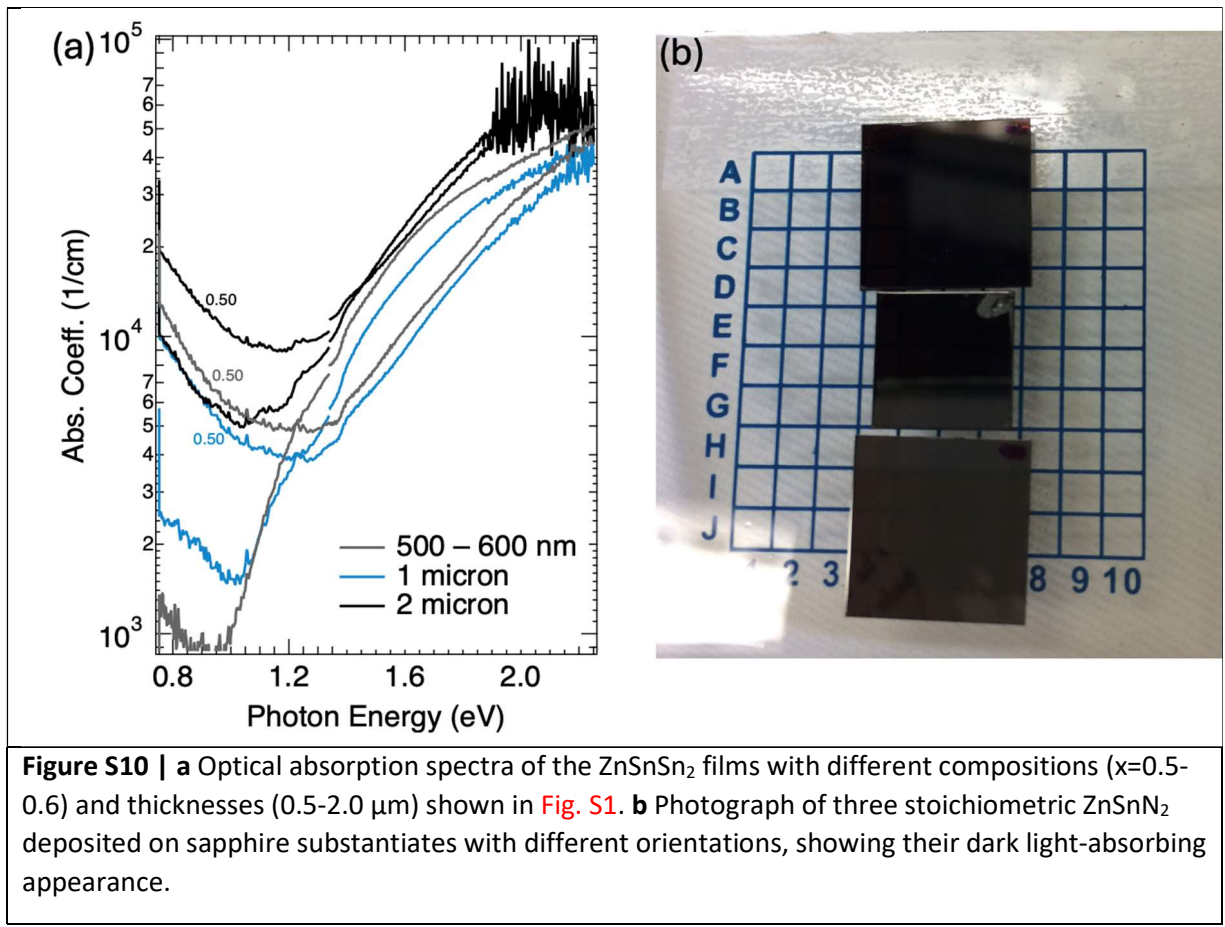

**Figure S10 | a** Optical absorption spectra of the ZnSnSn$_2$ films with different compositions (x=0.5-0.6) and thicknesses (0.5-2.0 μm) shown in Fig. S1. **b** Photograph of three stoichiometric ZnSnN$_2$ deposited on sapphire substantiates with different orientations, showing their dark light-absorbing appearance.